\documentclass[aps,prl,twocolumn,10pt,showpacs, citeautoscript,superscriptaddress]{revtex4-1} 

\usepackage[T1]{fontenc}	
\usepackage[english]{babel}	
\usepackage[latin1]{inputenc}	
\usepackage{times}

\usepackage[colorlinks=true, pdfstartview=FitV,linkcolor=blue, citecolor=blue, urlcolor=blue]{hyperref}

\usepackage{amsmath}		
\usepackage{amssymb}		
\usepackage{bbm}			

\usepackage{graphicx}		
\usepackage{graphics}		
\DeclareGraphicsExtensions{.png,.pdf}
\usepackage{color}		

\renewcommand{\theequation}{\arabic{equation}}
\newcommand{\Equation}[2]{\begin{equation}\label{#1}#2\end{equation}}
\newcommand{\Align}[2]{\begin{align}\label{#1}#2\end{align}}

\newcommand{\bs}{\boldsymbol}

\newcommand{\Figref}[1]{Fig.~\ref{#1}}

\newcommand{\Eqref}[1]{\eqref{#1}}

\newcommand{\Uone}{\mbox{U(1)}}
\newcommand{\Real}{\mathbbm{R}}

\newcommand{\dd}{\text{d}}

\newcommand{\Exp}[1]{\text{e}^{#1}}

\newcommand{\sg}{\text{sg}}
\renewcommand\Re{\mathrm{Re}}
\renewcommand\Im{\mathrm{Im}}
\newcommand{\grad}{\nabla}

\newcommand{\Lagrangian}{\mathcal{L}}
\newcommand{\PHI}{\mathbf{\Phi}}
\newcommand{\scale}{\bs{\eta}}

\newcommand{\munu}{{\mu\nu}}

\newcommand{\oa}{{(a)}}

\newcommand{\oo}{{(1)}}
\newcommand{\ot}{{(2)}}
\newcommand{\ga}{g_a}

\newcommand{\rmax}{r_{\mbox{\tiny max}}}

\begin{document}
\title{Stable Cosmic Vortons}
\author{Julien~Garaud}
\affiliation{
Department of Physics, University of Massachusetts Amherst, MA 01003 USA.
}
\affiliation{
Department of Theoretical Physics, The Royal Institute of Technology, Stockholm, SE-10691 SWEDEN.
}
\author{Eugen~Radu}
\affiliation{
Institut f\"ur Physik, Universit\"at Oldenburg, Postfach 2503 D-26111 Oldenburg, GERMANY.
}
\author{Mikhail~S.~Volkov}
\affiliation{
Laboratoire de Math\'{e}matiques et Physique Th\'{e}orique CNRS-UMR 7350, \\ 
Universit\'{e} de Tours, Parc de Grandmont, 37200 Tours, FRANCE.
}
\date{\today}

\begin{abstract}
We present for the first time solutions in the gauged 
$\Uone\times\Uone$ model of Witten describing vortons -- 
spinning flux loops stabilized against contraction by the 
centrifugal force. Vortons were heuristically described many 
years ago, however, the corresponding field theory solutions 
were not obtained and so the stability issue remained open. 
We construct explicitly a family of stationary vortons 
characterized by their charge and angular momentum. 
Most of them are unstable and break in pieces when perturbed. 
However, thick vortons with  small radius preserve their 
form in the $3+1$ non-linear dynamical evolution. 
This gives the first ever evidence of stable vortons and 
impacts several branches of physics where they could 
potentially exist. These range from cosmology, since 
vortons could perhaps contribute to dark matter, to QCD 
and condensed matter physics.
   
\end{abstract}

\maketitle


More than 25 years ago Witten introduced the idea of 
superconducting cosmic strings in the context of a field theory 
model that can be viewed as a sector of a Grand Unification Theory 
(GUT) \cite{Witten:85a}. The model admits classical solutions 
describing strings (vortices) whose longitudinal current can attain 
astronomical values (see \cite{vilenkin2000cosmic} for a review).

Soon after, it was realized that superconducting strings could form 
loops whose current would produce an angular momentum supporting 
them against contraction \cite{Davis.Shellard:89}. If stable, such 
cosmic {\sl vortons} should be of considerable physical interest, but 
until recently it was not clear if vortons are stable or 
not, since the underlying field theory solutions were not known.
Various approximations were used to describe vortons, for example, 
by viewing them as thin and large elastic rings \cite{Carter:1990sm}. 
It was also realized that objects similar to vortons could 
potentially exist also in other {domains}, as for example in 
condensed matter physics \cite{Battye.Cooper.ea:02,*Babaev:02b,
*Savage.Ruostekoski:03,*Metlitski.Zhitnitsky:04}, 
or in QCD \cite{Buckley:2002mx}. Since superconducting strings exist 
in the Weinberg-Salam theory \cite{Garaud.Volkov:10}, vortons are 
potentially possible also there. 

The first field theory solutions describing stationary vortons were 
found in the global limit of Witten's model, when the gauge fields 
vanish \cite{Radu.Volkov:08}. These vortons have approximately 
equal radius and thickness, like a Horn torus. Solutions describing 
thin and large vortons were later found as well, however, when perturbed, 
thin vortons turn out to be dynamically unstable and break in pieces 
\cite{Battye.Sutcliffe:09}. 
Although discouraging, this result is actually quite natural, since thin 
vortons can be locally approximated by straight strings, while the latter 
are known to become unstable for large currents \cite{vilenkin2000cosmic}. 

However, a more close inspection reveals that unstable modes of 
superconducting strings have a non-zero minimal wavelength 
\cite{Garaud.Volkov:10a,*Garaud.Volkov:08}, as in the case of 
the Plateau-Rayleigh instability of a water jet \cite{RevModPhys.69.865}. 
Therefore, imposing periodic boundary conditions with a short enough 
period should remove all instabilities. As a result, thick vortons 
made of short string pieces have chances to be stable.  

In this letter we present for the first time stationary vorton 
solutions in the gauged Witten's model, and our vortons are thick. 
To study their stability, we simulate their full $3+1$ non-linear 
dynamics in the limit of vanishing gauge couplings. We find that 
most of them are unstable, however, thick vortons with a large 
charge and the smallest possible radius are stable. By continuity, 
it follows that vortons with non-zero but small gauge couplings 
should be stable as well.      

We therefore present the first evidence of stable vortons, whose 
features turn out to be quite different from those predicted by the 
effective theories. This can impact several branches of physics 
where vortons could potentially exist.

{\bf The model of Witten.--}
\begin{figure*}[!htb]
  \hbox to \linewidth{ \hss
  \includegraphics[width=.25\linewidth]{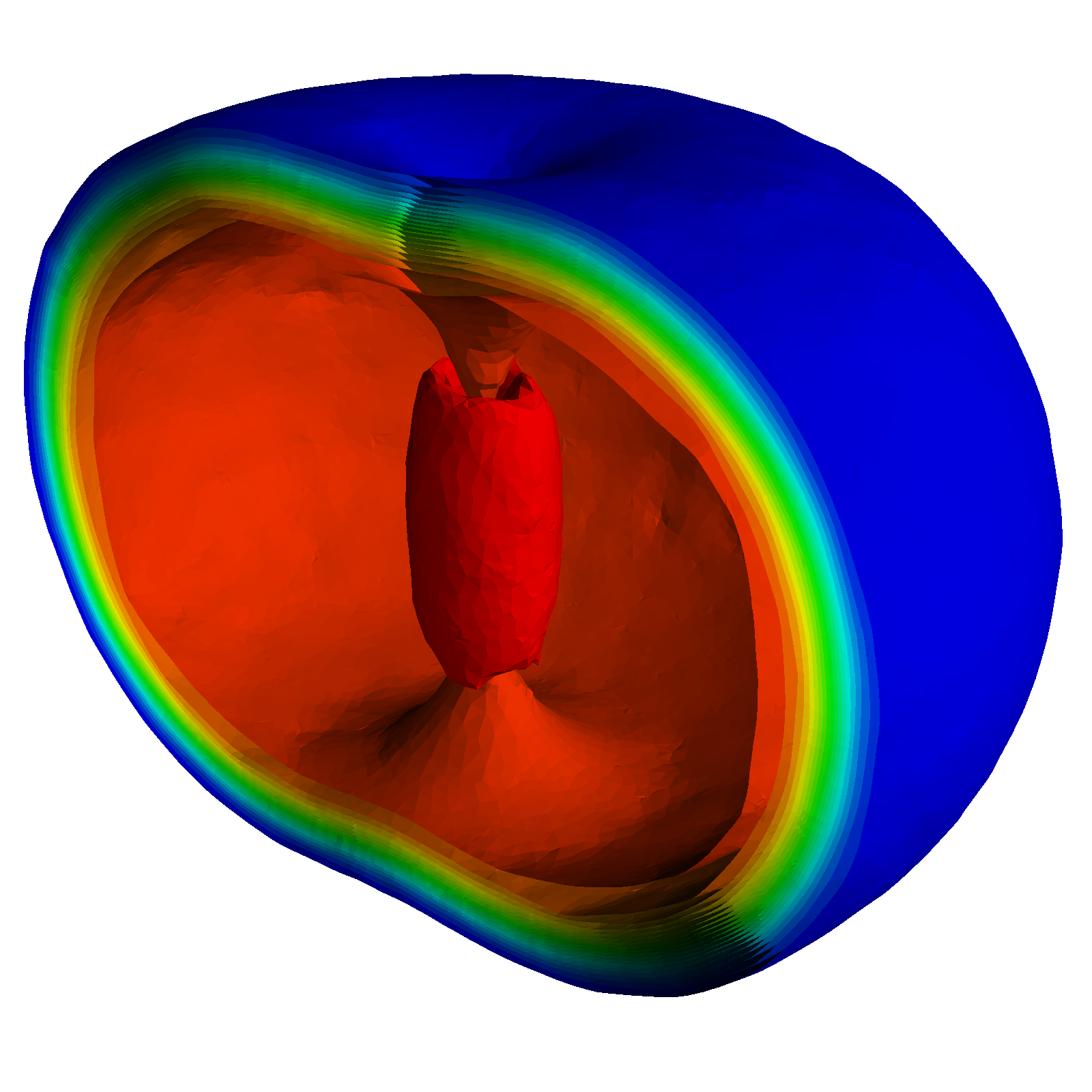}
  \includegraphics[width=.25\linewidth]{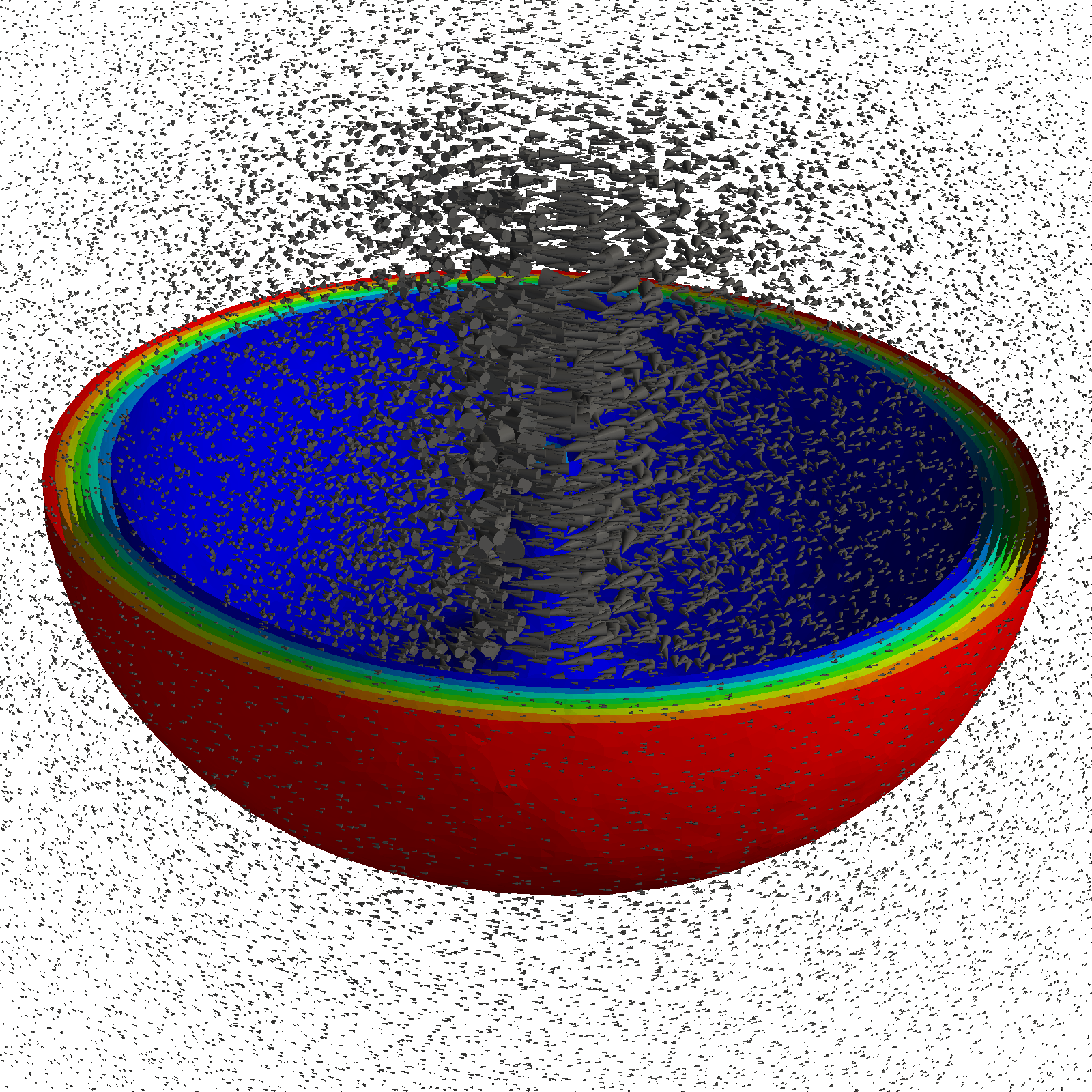}
  \includegraphics[width=.25\linewidth]{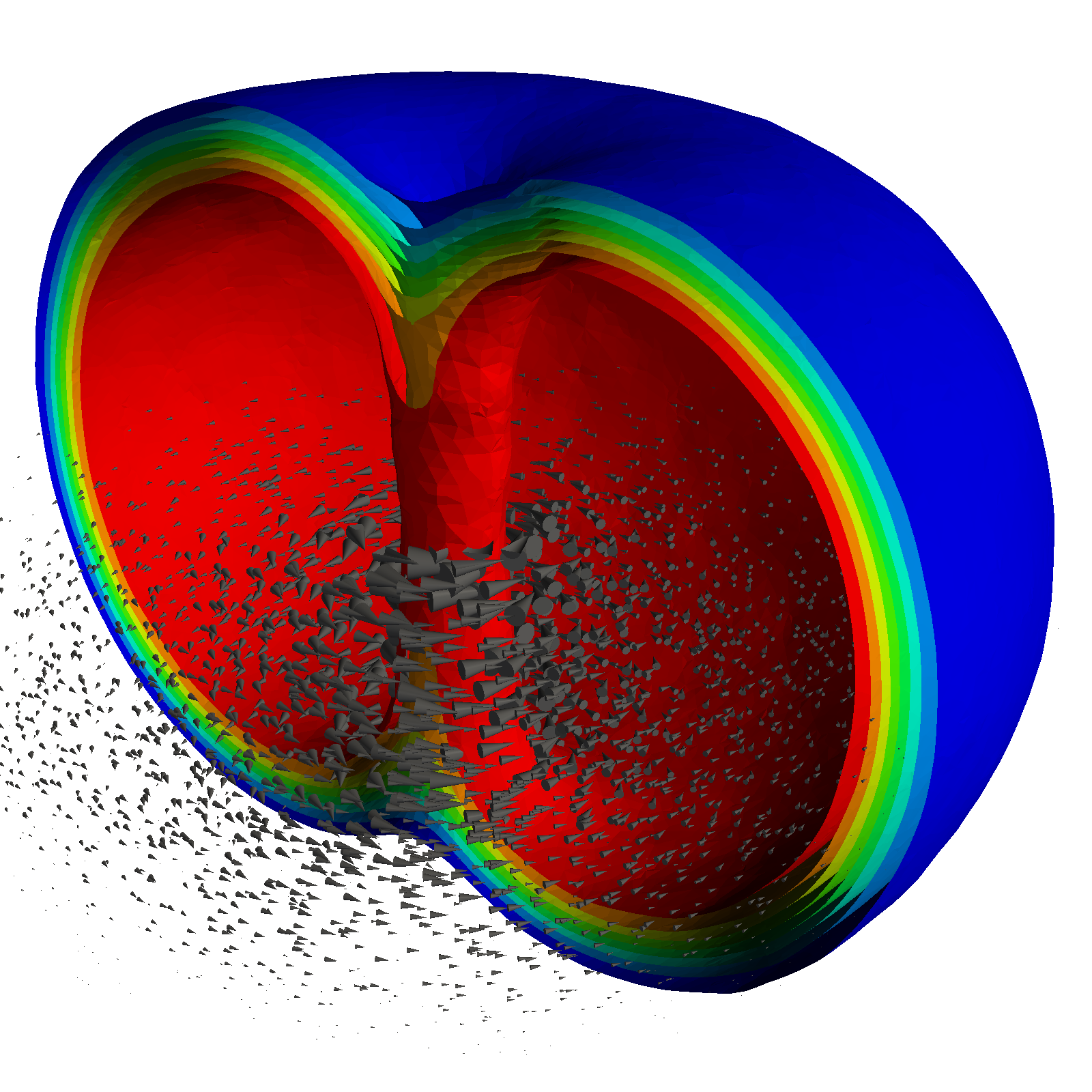}
  \includegraphics[width=.25\linewidth]{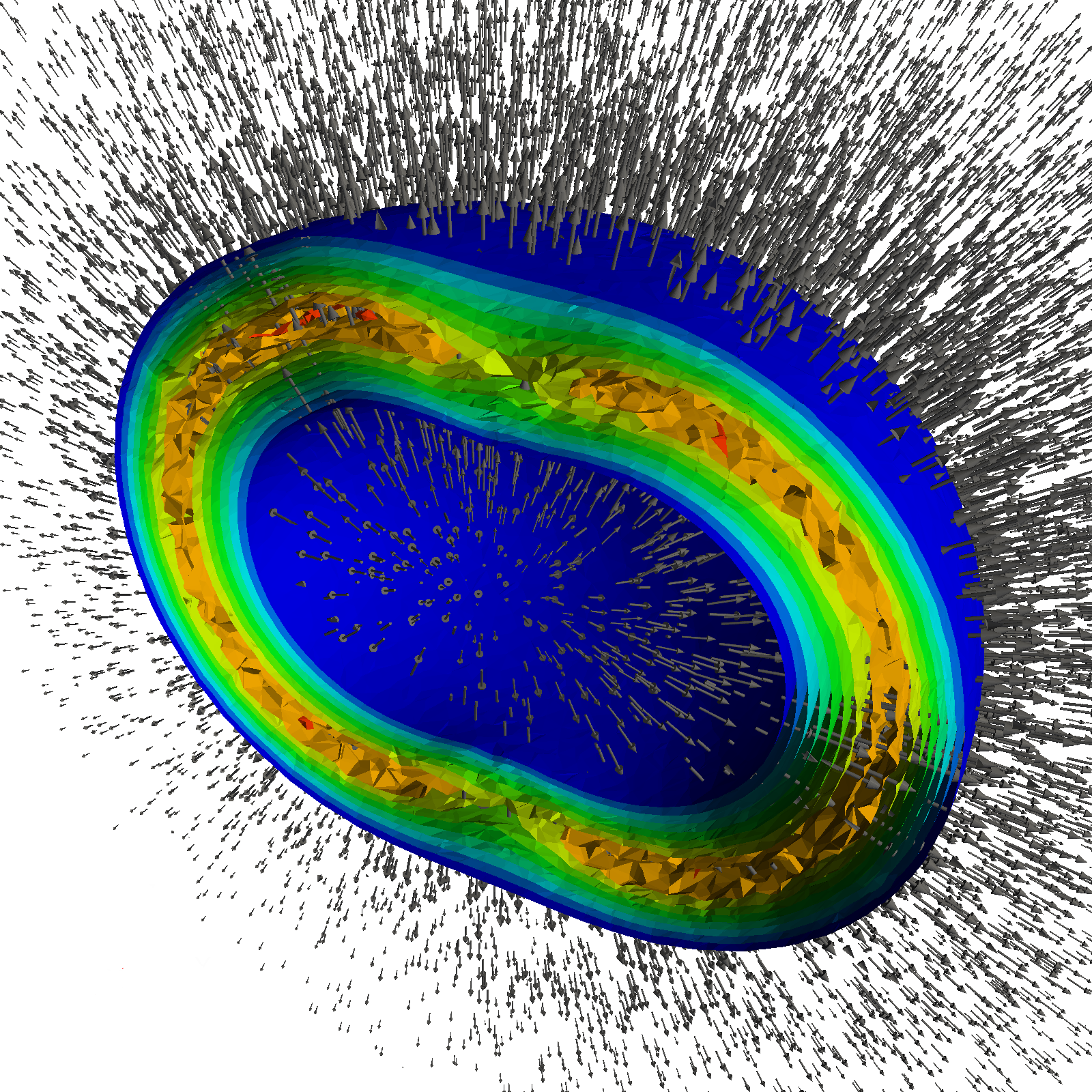}
  \hss}
\caption{
(Color online) -- 
Profiles of the stationary vorton solution for $Q=1500$ and $n=1$, 
$m=1$ for the parameter values $\lambda_1=41.1$, 
$(\lambda_2,\eta_2)=(30,1)$, $\gamma=20$, and $g_1=g_2=0.01$. 
The first panel displays constant energy surfaces. The second panel 
shows surfaces of constant $|\Phi_1|^2$ and the magnetic field 
$\vec{B}^\oo$ (cones). One has $\vec{E}^\oo=0$. The third panel 
shows $|\Phi_2|^2$ isosurfaces  and the electromagnetic current 
$\vec{j}_2$ (cones). The last panel shows the electric field 
$\vec{E}^\ot$ (cones), while the isosurfaces show its magnitude. 
The red color corresponds to large values and the blue color to 
small values.
}
\label{Fig:gauge-vorton}
\end{figure*}
This  is a theory of two Abelian vectors $A^{(a)}_\mu$ interacting 
with two complex scalars $\Phi_a$ $(a = 1, 2)$ with the Lagrangian 
\Equation{Action}{
{\cal L}=-\frac{1}{4} \sum_a F^\oa_\munu F^{\oa\,\munu} +
\sum_a(D_\mu\Phi_a)^*D^\mu\Phi_a
-V\,.
}
Here the gauge field strengths are
$F^{(a)}_{\mu\nu}=\partial_\mu A^{(a)}_\nu-\partial_\nu A^{(a)}_\nu$,
the gauge covariant derivatives 
$D_\mu\Phi_a=(\partial_\mu +ig_a A^\oa_{\mu} )\Phi_a$
with gauge couplings $g_a$, and the potential is
\Equation{Potential}{
V=\sum_a\frac{\lambda_a}{4}\left( |\Phi_a|^2-\eta^2_a\right)^2
	    +\gamma |\Phi_1|^2 |\Phi_2|^2 -\frac{\lambda_2\eta^4_2}{4}\,,
}
where $\eta_1=1$. If $4\gamma^2>\lambda_1\lambda_2$ and 
$2\gamma>\lambda_2\eta_2^2$ then the global minimum of the 
potential (vacuum) is achieved for $|\Phi_1|=1$ and $\Phi_2=0$. 
Fields $A^{(1)}_\mu$, $\Phi_1$, $\Phi_2$ are massive with masses, 
respectively, $m_A^2=2g_1^2$, $m_1^2=\lambda_1$, 
$m_2^2=\gamma-\frac12\lambda_2\eta_2^2$ whereas $A^{(2)}_\mu$ 
is massless and can be identified with electromagnetic field. 
The theory has a  local $\Uone\times\Uone$ invariance and two 
Noether currents, $j^{\mu}_a = 2\Re(i\Phi_a^\ast D^\mu\Phi_a)$, 
with two conserved charges $\int j^{t}_a\dd^3x$. The Euler-Lagrange 
equations are 
\Equation{EOM}{
\partial_\mu F^{(a)\mu\nu}= g_a j_a^\nu\,,~~~~~
D_\mu D^\mu \Phi_a +\frac{\partial V}{\partial|\Phi_a|^2}\Phi_a=0\,.
}
Assuming cylindrical coordinates  $x^\mu=(t,\rho,z,\varphi)$, 
we make the ansatz for stationary, axially symmetric fields, 
\Align{fields}{
   \Phi_1 &= X_1+iY_1\,, 
   \Phi_2 = (X_2+iY_2)\exp\{i(\omega t+m\varphi) \}\,, 
}
where $X_a$, $Y_a$ as well as $A^{(a)}_\mu$ depend on $\rho,z$, 
and we impose the gauge condition $A^{(a)}_\rho=0$. Here $m$ is 
an integer winding number and $\omega$ is a frequency. The fields 
should be globally regular and the energy should be finite, which 
requires that at infinity $X_1\to 1$ while all other amplitudes 
approach zero. At the symmetry axis, $\rho=0$, the amplitudes 
$X_2,Y_2,A^{(a)}_\varphi$ vanish, while for the other amplitudes 
the normal derivative $\partial/\partial\rho$ vanishes. Under 
the reflection $z\to-z$ the amplitudes $Y_a$ are odd whereas 
all the others are even.  

The choice of the ansatz implies that the first Noether charge vanishes, 
while the second one is
\Equation{Q}{
Q =	2\int \dd^3x\  \left(X_2^2+Y_2^2\right)\left(\omega-g_2 A^{(2)}_t\right).
}
The energy is $E=\int T^t_t d^3x$ and the angular momentum 
\Equation{AngMom}{
   J=\int T^t_\varphi\dd^3x=mQ,
}
where the energy-momentum tensor is obtained by varying the metric 
tensor, $T^\mu_{~\nu}=2g^{\mu\sigma}\partial{\cal L}/\partial g^{\sigma\nu}
-\delta^\mu_{\nu}\,{\cal L}$. In the above formulas all fields and 
coordinates are dimensionless. If $\scale$ is the energy scale, then 
the dimensionful (boldfaced) quantities are $\PHI_a=\scale\Phi_a$, 
${\bf A}^{(a)}_\mu=\scale A^{(a)}_\mu$, $x^\mu={\bf x}^\mu\scale$, 
${\bf E}=\scale E$, hence $\scale$ is the asymptotic value of $\PHI_1$. 
 
{\bf Stationary vortons.--}
Inserting the ansatz \eqref{fields} to the field equations \Eqref{EOM} 
gives, after separating  $t$ and $\varphi$ variables, an elliptic system 
of $10$ non-linear PDE's for the $10$ functions of $\rho,z$. We solve 
these equations with two different numerical methods: using the elliptic 
PDE solver FIDISOL based on the Newton-Raphson procedure 
\cite{Schonauer.Schnepf:87}, and also minimizing the energy within 
a finite element approach provided by the Freefem++ library \cite{Freefem}.

We look for solutions with a toroidal structure and non-trivial phase 
windings along both torus generators. Apart from the azimuthal winding 
number $m$, there is a second integer, $n$, determining the winding 
of phase of $\Phi_1$ around the boundary of the $(\rho,z)$ half-plane. 
If $n\neq 0$ then $\Phi_1$ vanishes at a point $(\rho_0,0)$ 
corresponding to the center of the closed vortex forming the vorton, 
and the phase of $\Phi_1$ winds around this point. Prescribing non-zero 
values of $n,m$, and $Q$ (see \cite{supplementary} for details),
the fields cannot unwind to vacuum, and the iterative numerical procedure 
converges to a smooth limiting configuration with a finite radius $\rho_0$. 
We have constructed in this way vortons for $n = 1,2$ and  
$m = 1,\ldots, 12$, and also solutions similar to Q-balls 
\cite{Lee.Stein-Schabes.ea:89} for $n=0$, $m=0,1,\ldots $ 
\cite{supplementary}.

The vorton  can be visualized as  a toroidal tube confining a 
magnetic flux of $\vec{B}^{(1)}=\vec{\nabla}\times\vec{A}^{(1)}$, 
since $\Phi_1\approx 0$ inside the tube  and thus the first $\Uone$ 
is restored.  $\Phi_2$ is non-zero inside the tube, giving rise to 
a charged condensate and to a persistent electric current along 
the tube. The current creates a momentum along the azimuthal 
direction, which gives rise to an angular momentum along $z$-direction. 
Outside the vorton tube the massive fields $A^{(1)}_\mu$, $\Phi_1$, 
$\Phi_2$ rapidly approach their vacuum values and there remains 
only the long-range massless $A^{(2)}_\mu$ generated by the electric 
current confined inside the vorton tube. At large $r=\sqrt{\rho^2+z^2}$ 
one has $A^{(2)}_t={\cal Q}/(4\pi r)+\ldots $ and 
$A^{(2)}_\varphi =\mu \sin \theta/r^2 +\dots $, therefore, from far 
away the vorton looks like a superposition of an electric charge 
${\cal Q}$ with a magnetic dipole $\mu$. 

\Figref{Fig:gauge-vorton} shows the 3D solution profiles for an 
$m=1$ vorton. One can see that the vorton tube is very thick and 
compact. The vortex magnetic field 
$\vec{B}^{(1)}=\vec{\nabla}\times\vec{A}^{(1)}$ and the electric 
current $\vec{j}_2$ are tangent to the azimuthal lines. The electric 
field $\vec{E}^{(2)}=-\vec{\nabla}{A}^{(2)}_t$ is mostly oriented 
along the radial direction and supports a non-zero flux at infinity, 
${\cal Q}=\oint \dd\vec{E}^{(2)}\cdot \dd\vec{S}=g_2Q$. The massless 
magnetic field $\vec{B}^{(2)}=\vec{\nabla}\times\vec{A}^{(2)}$ at 
large $r$ is of magnetic dipole type. 

Vortons can be labelled by their charge $Q$ and the integer `spin' 
$m=J/Q$ (assuming that $n=1$). The vorton radius $\rho_0$ is not 
very sensitive to the value of $Q$ but increases rapidly with $m$, 
so that for large $m$ vortons are thin and large, with the radius 
much larger than the thickness. On the other hand, increasing $Q$ 
increases the thickness of the vorton tube, so that for large $Q$ 
vortons are thick and look almost spherical. 

\begin{figure}[!htb]
   \hbox to \linewidth{ \hss
   \includegraphics[width=0.5\linewidth]{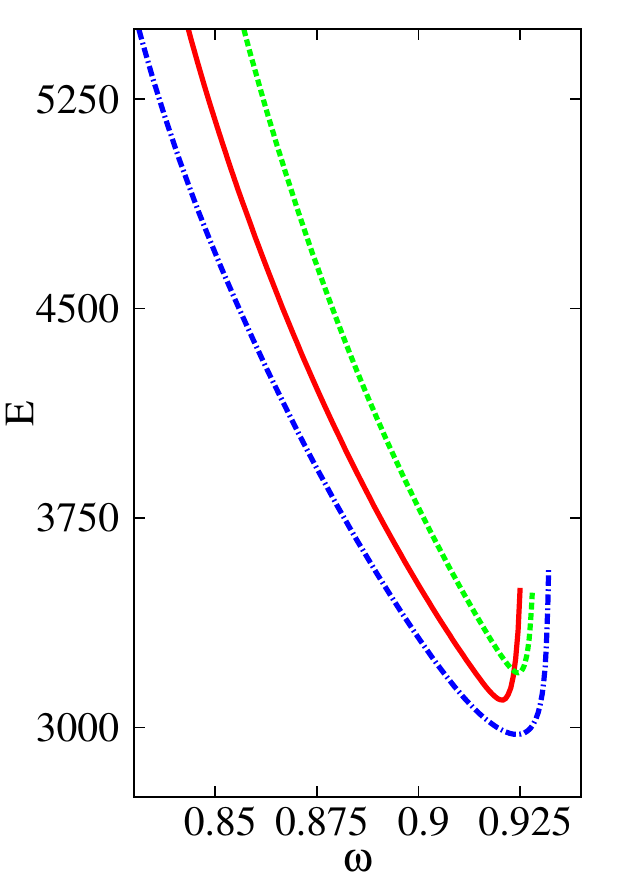}
   \includegraphics[width=0.5\linewidth]{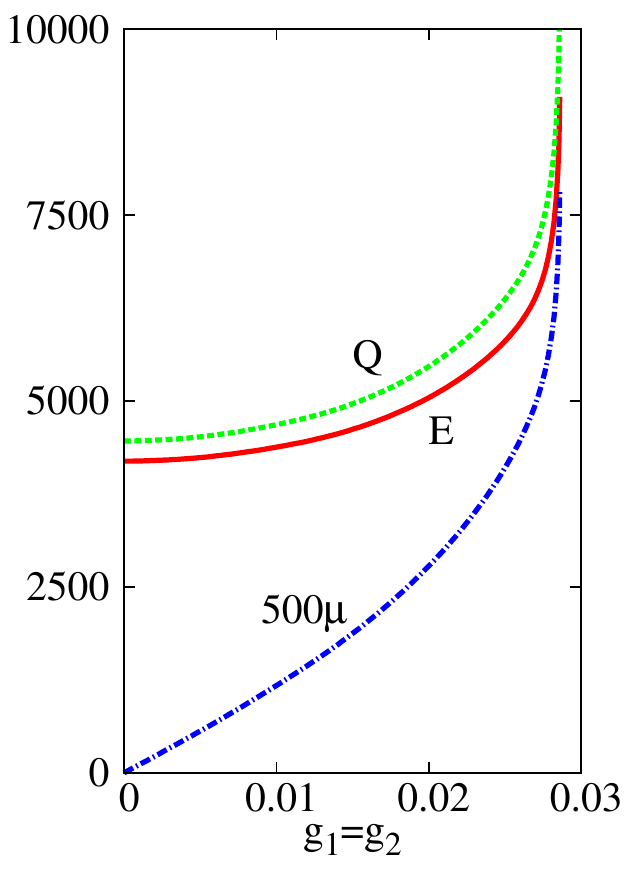}
   \hss}
\caption{(Color online) -- 
Left panel shows $E$ against $\omega$ for the values of gauge 
couplings $(g_1,g_2)=(0.008,0.021)$ (dashed), $(0.012,0.012)$ 
(solid), $(0.008,0.027)$ {(dotted)}. Right panel shows $E$, $Q$, 
$\mu$ against $g_1=g_2$ for a fixed $\omega=0.804$. In both cases 
$\lambda_1=41.1$, $\lambda_2=40$, $\gamma=22.3$, $\eta_2=1$, 
$n=1$, $m=2$. 
}
\label{Fig:curve1}
\end{figure}

The frequency $\omega$ can be used instead of $Q$ to characterize 
the solutions, which exist only within a finite frequency range, 
$\omega_{-}<\omega<\omega_{+}$. Both $E$ and $Q$ diverge for 
$\omega\to \omega_\pm$ and have a minimum in between, as shown in 
\Figref{Fig:curve1}. Vortons can be viewed as boson condensates, which 
is why their charge cannot be too small, since the boson condensation 
is not energetically favoured for small quantities of the field quanta. 

For $g_a=0$ the gauge fields vanish and the vortons are global, 
`made of' the scalars $\Phi_a$ alone  \cite{Radu.Volkov:08}. For 
$g_a\neq 0$ the gauge fields are excited and increase the total 
energy and charge, as shown in \Figref{Fig:curve1}. Solutions do 
not exist for arbitrary values of $\lambda_a$, $\eta_2$, $\gamma$, 
$g_a$  but only for some regions in the parameter space. For example, 
fixing all parameters and also $\omega$ and varying $g_1=g_2$, the 
solutions exist only within a finite range of gauge couplings, as is 
seen in \Figref{Fig:curve1}.

{\bf Dynamical vortons.--}
To analyze the vorton stability, we simulate their non-linear 
$3+1$ temporal dynamics. In doing this, we consider only the global 
vortons, since simulating dynamics of the gauge fields would require 
too much computer power. However, we expect the results obtained in 
the global case to apply to the fully gauged vortons as well, at least 
for small enough gauge couplings $g_a$. Indeed, for $g_a=0$ the vorton 
is made of scalars $\Phi_a$. For small nonzero $g_a$ their global 
currents give rise to the $O(g_a)$ source in the gauge field equations, 
hence $A_\mu^{(a)}=O(g_a)$. The backreaction of the gauge fields on the 
scalars is  $O(g_a^2)$ and can be neglected as compared to the reaction 
of the scalars on themselves. For stationary vortons this is confirmed 
by the numerics, as is seen in \Figref{Fig:curve1}. Therefore, one can 
expect the temporal dynamics of vortons with small $g_a$ to be dominated 
by the scalars only, hence it can be approximated by the global vorton 
dynamics.

\begin{figure}[!htb]
 \hbox to \linewidth{ \hss
  \includegraphics[width=.25\linewidth]{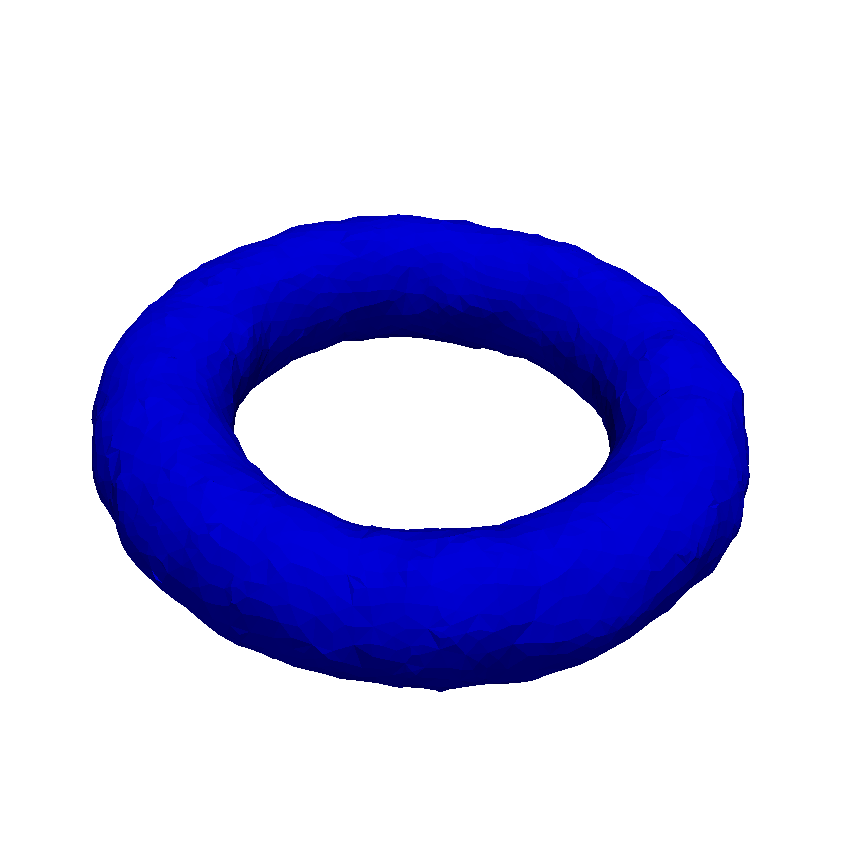}
  \includegraphics[width=.25\linewidth]{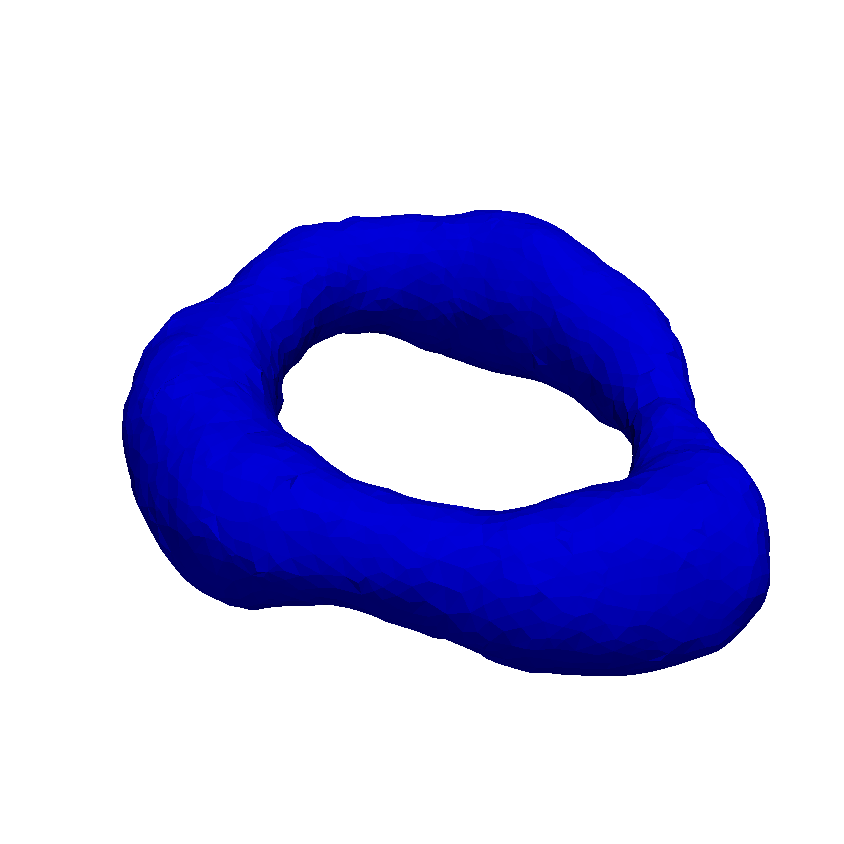}
  \includegraphics[width=.25\linewidth]{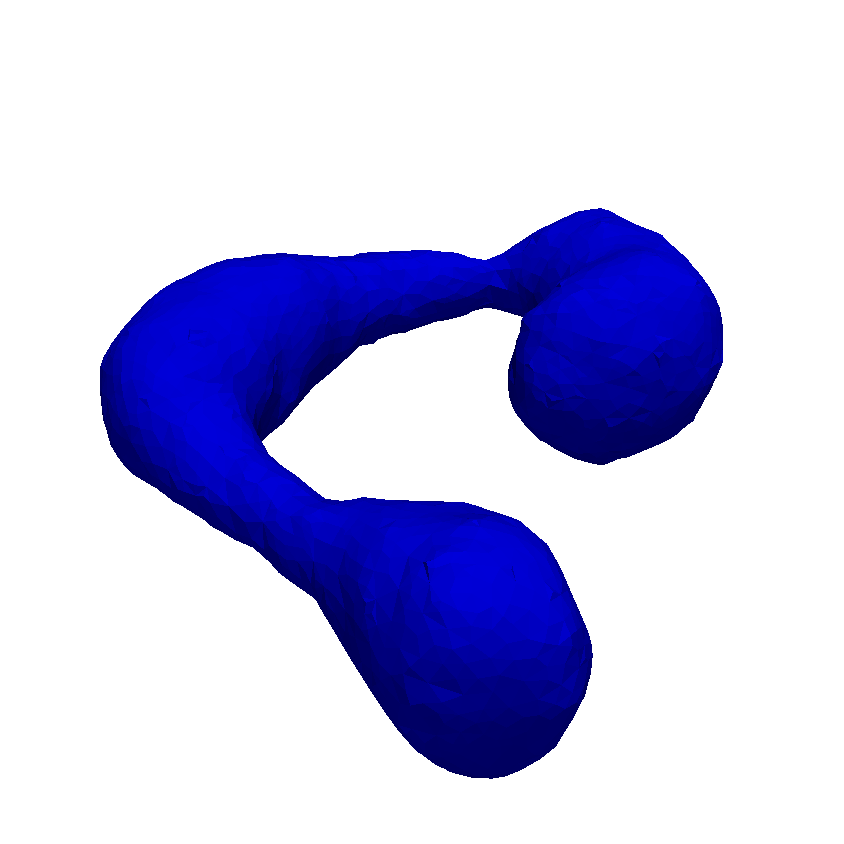}
  \includegraphics[width=.25\linewidth]{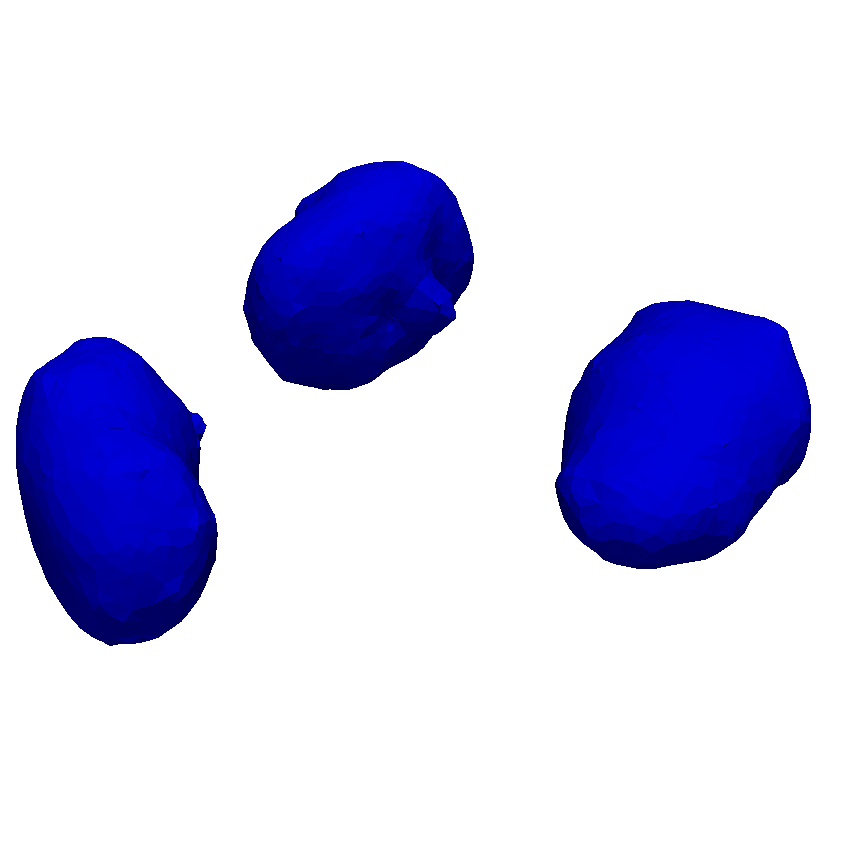}
  \hss}
 \hbox to \linewidth{ \hss
  \includegraphics[width=.25\linewidth]{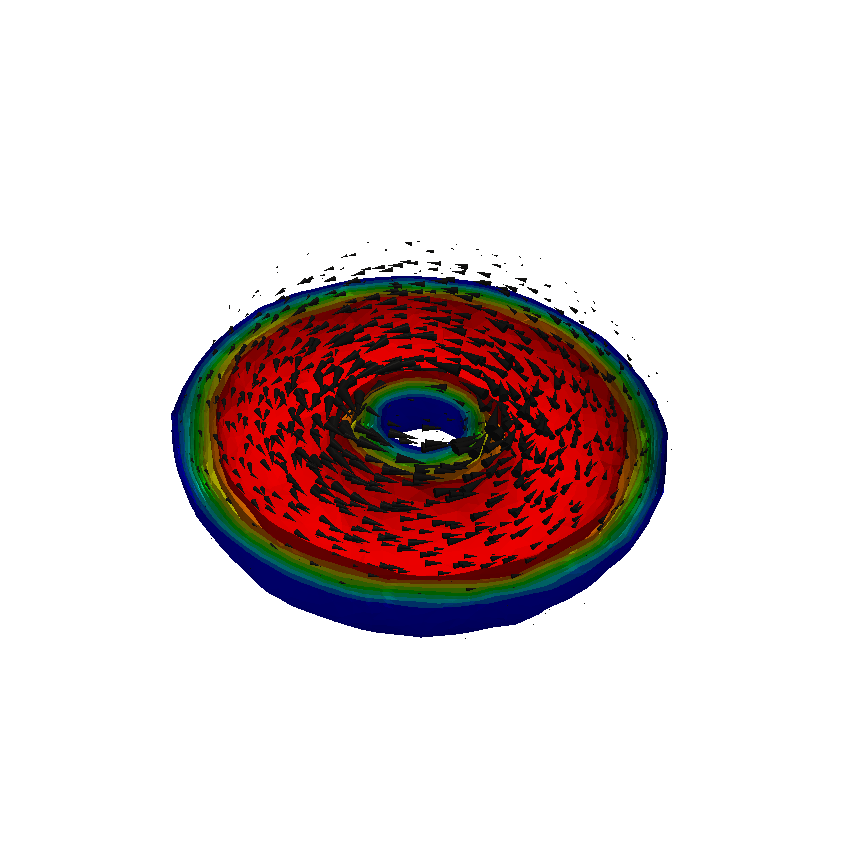}
  \includegraphics[width=.25\linewidth]{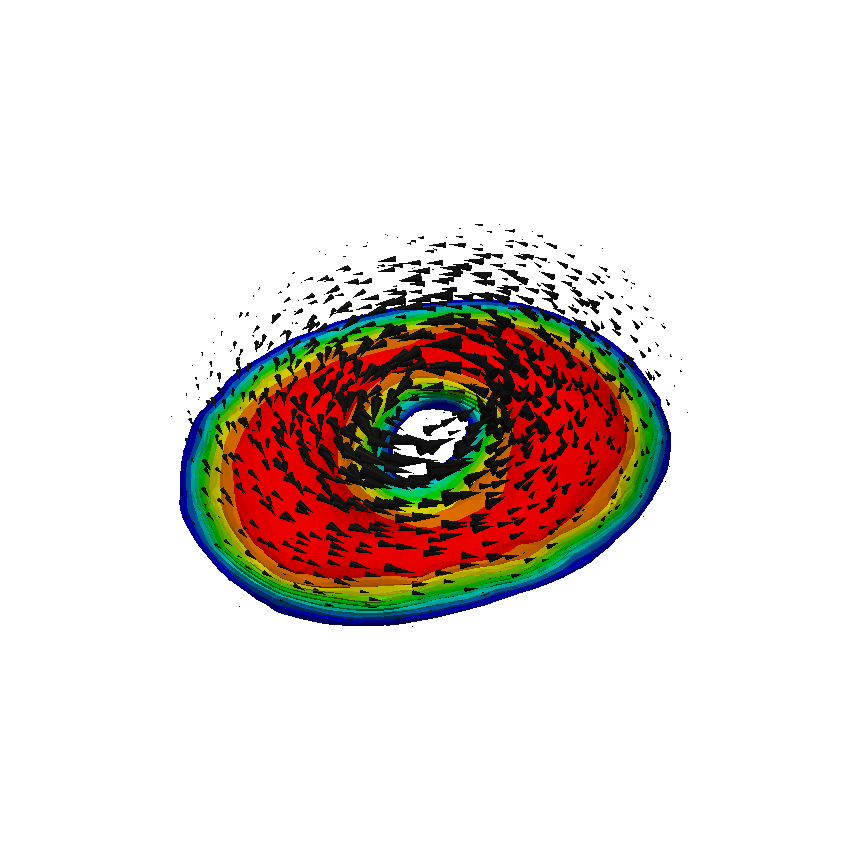}
  \includegraphics[width=.25\linewidth]{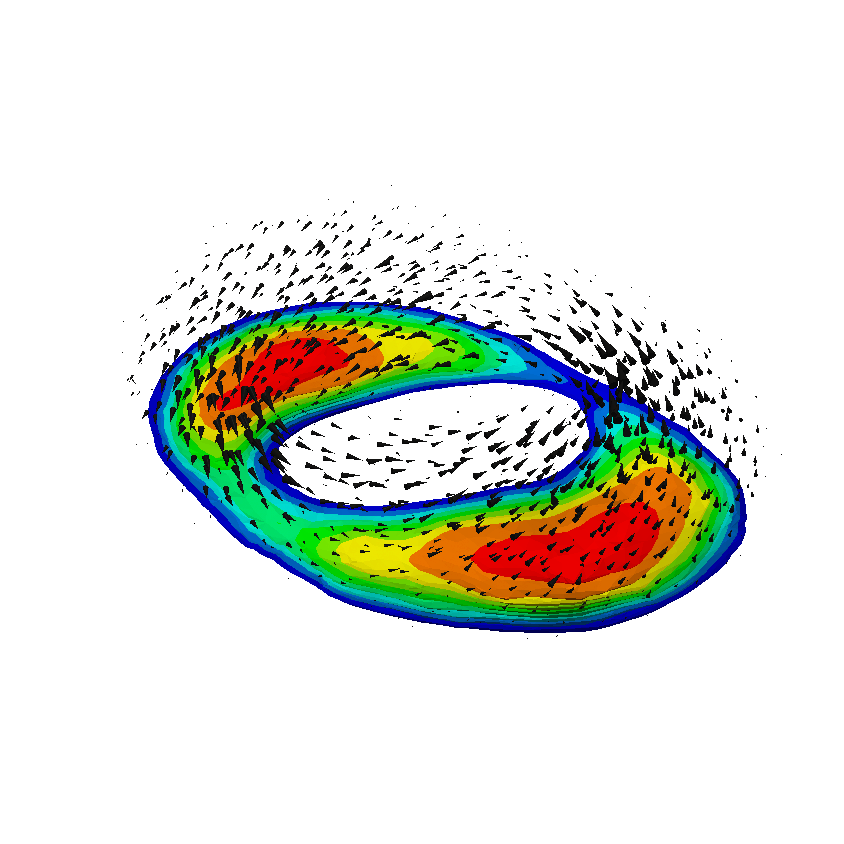}
  \includegraphics[width=.25\linewidth]{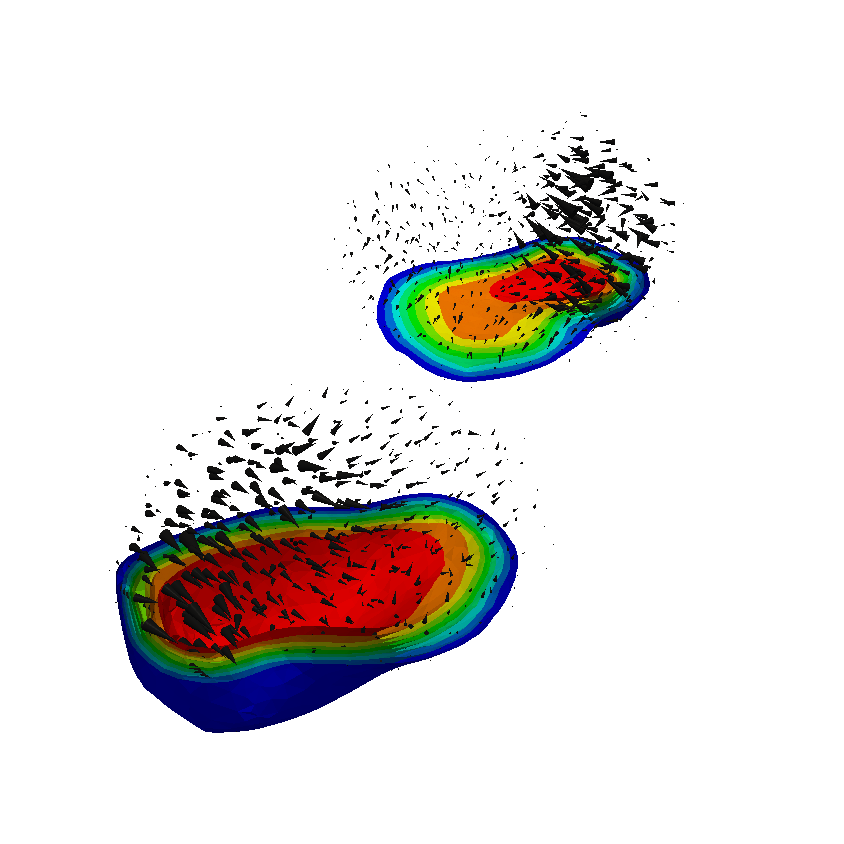}
  \hss}
 \hbox to \linewidth{ \hss
  \includegraphics[width=.25\linewidth]{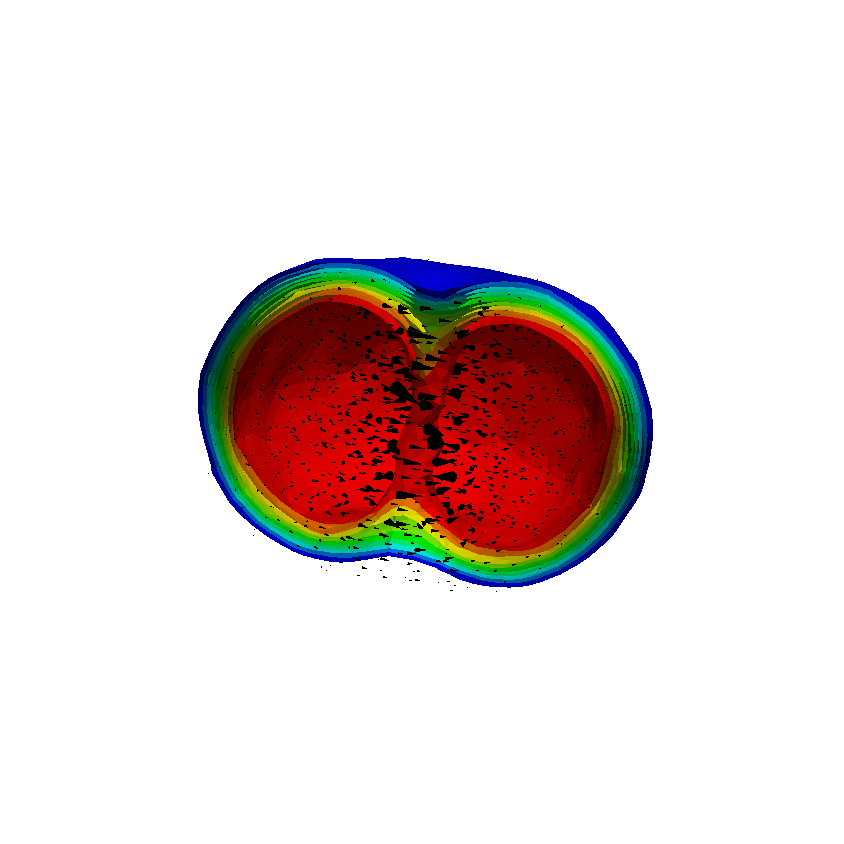}
  \includegraphics[width=.25\linewidth]{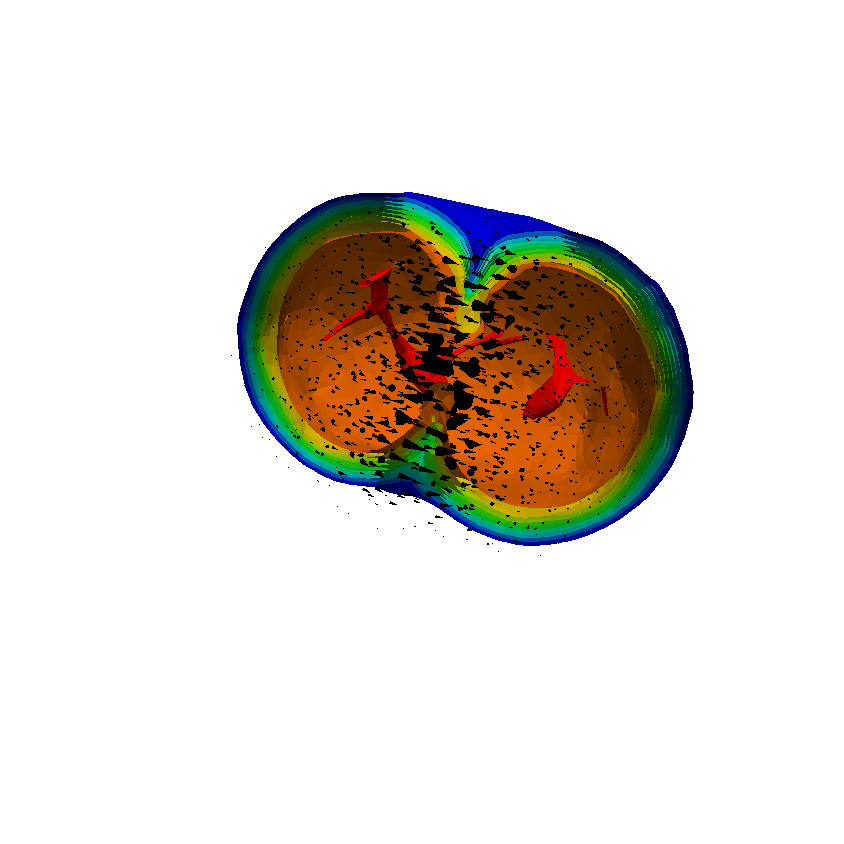}
  \includegraphics[width=.25\linewidth]{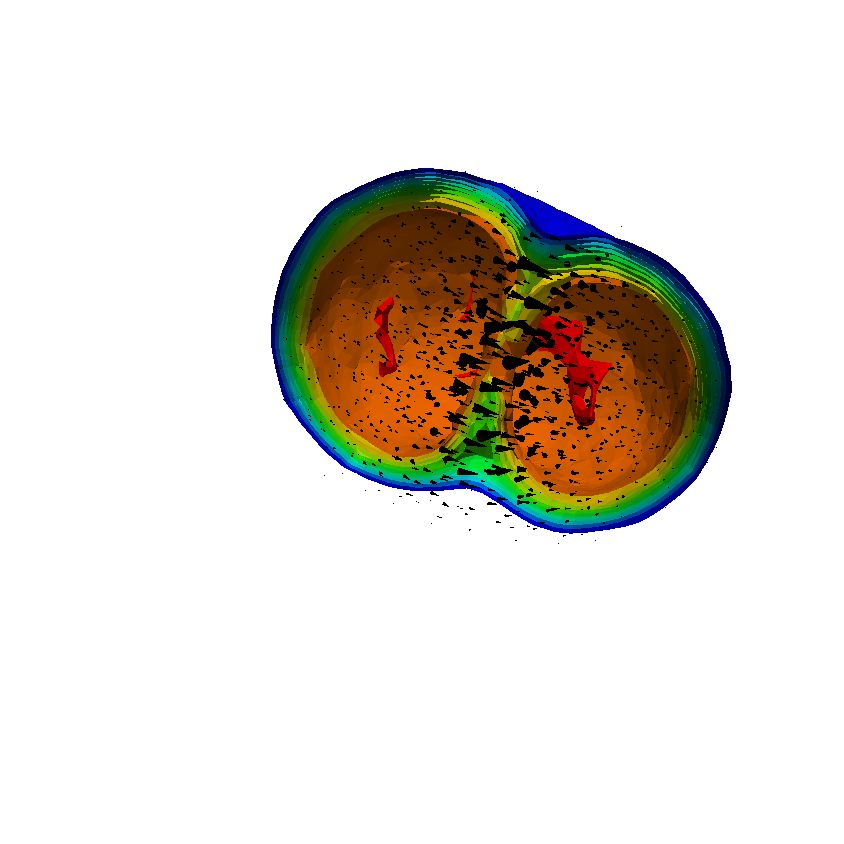}
  \includegraphics[width=.25\linewidth]{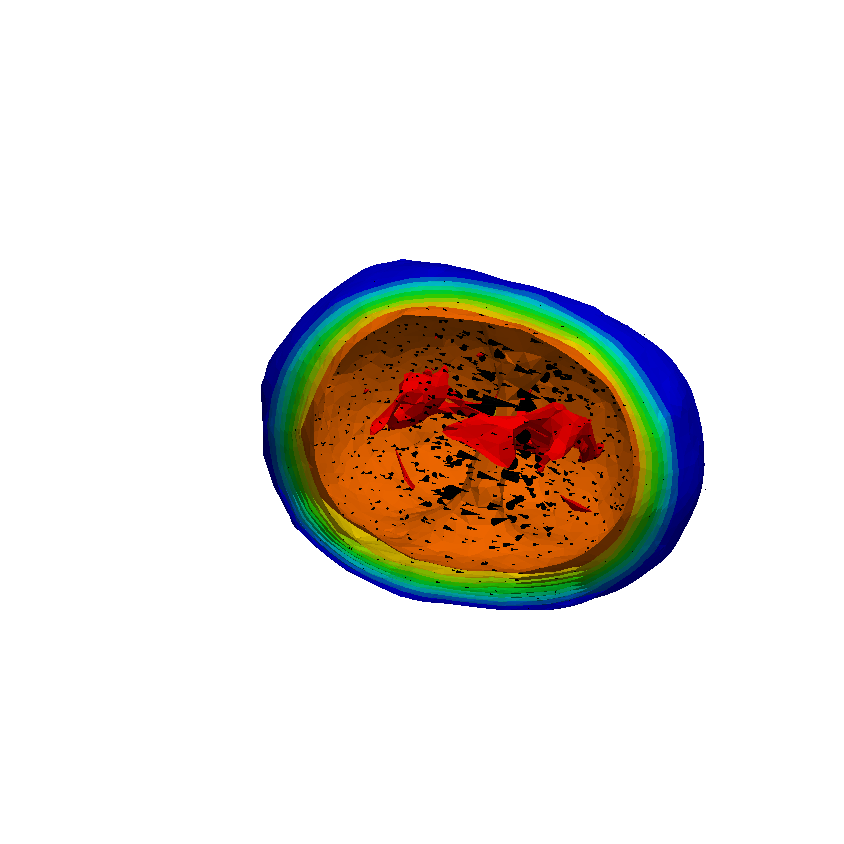}
  \hss}

\caption{
(Color online) -- 
Snapshots of the vorton time evolution. The first line shows a 
constant $|\Phi_2|^2$ surface for the $m=6$ solution for $t=0,17,21,28$. 
The second and third lines show the constant $|\Phi_1|^2$ surfaces and 
the electric current, respectively, for $m=3$ vorton for $t=0,31,28,47$ 
and for the $m=1$ vorton for $t=0,104,153,239$. The third line shows 
the stable solution for $Q=6000$ and for the same values of $\lambda_a$, 
$\eta_2$,$\gamma$ as in \Figref{Fig:curve1}. 
}
\label{Fig:dyn1}
\end{figure}

Vortons with large $m$ develop  pinching deformations breaking 
them in pieces \cite{Battye.Sutcliffe:09}. However, for small $m$, 
vortons could be stable, since they are compact and thick and have 
no room for the instability to settle in. To verify this, we consider 
a hyperbolic evolution scheme based on an implicit $\beta$-Newmark 
finite difference approximation (see \cite{supplementary} for details).
The initial configuration is a stationary, axially symmetric vorton.
It becomes automatically perturbed by the space discretization when 
adapted from 2D to the 3D mesh, which triggers a non-trivial temporal 
evolution. The natural timescale is set by the value of $\omega$ of 
the underlying vorton solution, which is of order one.
We integrate with the timestep $\Delta t=0.1$ and find that the 
$m\geq 3$ vortons very quickly become strongly deformed and then break 
in pieces. The time they take to break decreases rapidly as $m$ grows 
(see \Figref{Fig:dyn1} and \cite{supplementary} for the videos). 
The products of the vorton decay are typically two or three out-spiralling 
fragments of spherical topology.

We therefore conclude that thin and large vortons are unstable, thus 
confirming the result of \cite{Battye.Sutcliffe:09}. One should say 
that a different conclusion was previously made in 
Ref.~\cite{Lemperiere.Shellard:03a}, where thin vortons were found 
to be stable. Since neither our analysis nor that of 
\cite{Battye.Sutcliffe:09} confirm this, it is possible that the 
conclusion of \cite{Lemperiere.Shellard:03a} is an artefact of 
modifying the scalar potential made in that work in order to improve 
the stability behavior.  

We finally turn to vortons with $m=1,2$ and choose a large value $Q$,
in which case the vortons are compact and thick. For $m=2$ we cannot 
make a definite conclusion, since the vortons do not actually break but 
sometimes become strongly deformed.  
However, nothing at all happens to the $m=1$ vortons. As the time goes, 
they only move slowly  in the box, sometimes reflecting from the boundary, 
without changing shape. We integrated up to $t\sim 10^3$ (which requires 
weeks of runtime) without noticing any change in their behavior. 
We also checked that increasing the size of the box does not change 
anything, so that one cannot say that the boundary has a stabilizing 
effect. We therefore conclude that vortons with the lowest `spin' and 
a large charge are dynamically stable (see \cite{supplementary} for 
more discussion). Intuitively, this is because they are so thick that 
they are `hard to pinch'.   

Interestingly,  a very similar conclusion was made for the 
`spinning light bullets' which share many properties with vortons 
\cite{PhysRevLett.88.073902}. These are non-relativistic solutions 
for a complex scalar field with a $t,\varphi$-dependent phase (like 
for $\Phi_2$), and they also have toroidal profiles. Their $E(\omega)$ 
dependence   is similar to that shown in \Figref{Fig:curve1}. It was 
found that the $m=1$ solutions with a large charge  and $\omega$ 
close to $\omega_{-}$ preserve their shape in the $3+1$ dynamical 
evolution \cite{PhysRevLett.88.073902}. Exactly the same statement 
applies for our relativistic vortons. 

In summary, for the first time since the vortons were heuristically 
described almost 25 year ago \cite{Davis.Shellard:89}, we present 
the underlying stable solutions within the $\Uone\times\Uone$ gauge 
field theory of Witten \cite{Witten:85a}. We can now make some 
estimates. Assuming the original motivation of Witten, the energy 
scale should be of the GUT magnitude, $\scale\sim 10^{14}$ GeV. 
Using the value $E\sim 5\times 10^{3}$ for estimates, it follows 
that vortons are extremely heavy, with ${\bf E}\sim 5\times 10^{17}$ GeV,
which is not far from the Planck energy. On the other hand, 
their minimal Noether charge $Q\sim 5\times 10^{3}$ is actually 
not so large as compared to the average particle density in the hot 
early universe. Therefore, vortons could be abundantly created due 
to charge fluctuations in the course of a phase transition via the 
Kibble mechanism \cite{PhysRevLett.60.2101,*PhysRevD.40.3529},
if only GUTs indeed applied in the past. Being classically stable, 
vortons could disintegrate via a quantum tunnelling towards the 
$\rho_0=0$ state, but this process should be exponentially suppressed, 
and in fact quantum fluctuations can also have a stabilizing effect 
by preventing the collapse to zero size \cite{Graham}. Vortons could 
probably evaporate via interactions with GUT fermions 
\cite{Kusenko:1997si}, but this process should stop after the GUT epoch.
Therefore, it is not inconceivable that some relic vortons could still 
be around and contribute to dark matter. 

Let us consider $g_a=0$ vortons. For stationary fields \eqref{fields}, 
Eqs.\Eqref{EOM} for $\Phi_a$ can formally be interpreted 
\cite{Radu.Volkov:08} as the non-relativistic Gross-Pitaevskii 
equation for a two-component Bose-Einstein condensate (BEC). This 
can describe ultracold atomic gazes with two hyperfine states, 
such as $^{87}$Rb \cite{Myatt:1997zz,*PhysRevLett.81.1543}.
Scalars $\Phi_a$ then correspond to the two BEC order parameters, 
one of which creates a vortex while the other one condenses in the 
vortex core. Our solutions therefore describe stationary vortons 
made of loops of such vortices, whose potential existence has been 
much discussed \cite{Battye.Cooper.ea:02,*Babaev:02b,
*Savage.Ruostekoski:03,*Metlitski.Zhitnitsky:04}.
In fact, vortex rings in two-component BECs have been observed  
experimentally \cite{PhysRevLett.86.2926}, although it is not 
completely clear if they support an angular momentum 
\cite{Nitta.Kasamatsu.ea:12}.

$\Phi_1$ and $\Phi_2$ can also be interpreted \cite{Buckley:2002pj}, 
respectively, as the $d$-wave superconducting (dSC) and antiferromagnetic 
(AF) order parameters in the SO(5) model of high $T_c$ superconductivity 
\cite{Zhang}. This model admits dSC  vortices with an AF core \cite{PhysRevLett.79.2871,*PhysRevB.60.6901}, while our solutions describe 
loops made of these vortices. Such vorton quasiparticles could be important 
for the  superconducting phase transition in this model \cite{Buckley:2002pj}.

Equally, scalars $\Phi_a$ can be viewed as describing a condensate 
of $(K^{+},K^0)$ mesons in QCD \cite{Kaplan:2001hh}, hence our 
solutions describe the $K$-vortons, whose existence was conjectured 
in \cite{Buckley:2002mx}. Setting the scale to be $\scale\sim 200$ MeV 
gives for their energy ${\bf E}\sim 1$ TeV. Such objects could probably 
exist in dense QCD matter, as for example in neutron stars 
\cite{Buckley:2003zf,*PhysRevC.69.055803}, which may affect their 
electromagnetic and neutrino transport properties.

\acknowledgments
J.G. was supported by the NSF grant No. DMR-0955902 and by 
the Swedish Research Council.
A part of this work 
was performed at the Royal Institute of Technology on resources 
provided by the Swedish National Infrastructure for Computing  at the 
National Supercomputer Center in Linkoping, Sweden.
E.R. gratefully acknowledges support by the DFG.

\bibliographystyle{apsrev4-1}

\section{Appendix} \label{Appendix1}
\renewcommand{\theequation}{A.\arabic{equation}}
\setcounter{equation}{0}

We used three completely different numerical techniques for the 
vorton construction. First, we performed our calculations using 
the elliptic PDE solver FIDISOL based on the iterative Newton-Raphson 
method \cite{Schonauer.Schnepf:87} within a finite difference scheme. 
This method solves the equations of motion for the stationary, 
axially symmetric fields. To construct a solution with this method, 
one fixes the values of the winding numbers $n,m$ and also the 
frequency $\omega$ in the ansatz \Eqref{fields2}, whereas $Q$ 
is obtained from Eq.\Eqref{QQ}. This method requires a good choice 
of the starting field configuration. 
Next, we reproduced the same solutions using the energy minimization 
scheme within a finite element framework provided by the Freefem++ 
library \cite{Freefem}. In this case the input parameters are 
$n,m,Q$, while $\omega$ is obtained from \eqref{omega}. This method 
is less sensitive to the  choice of the input field configuration.
Finally, we simulated the hyperbolic evolution of the full $3+1$ theory, 
also using the Freefem++ library \cite{Freefem}. Below we shall describe 
the two last methods.  

\subsection{Energy minimization}

\begin{figure}[!htb]
  \hbox to \linewidth{ \hss
   \includegraphics[width=\linewidth]{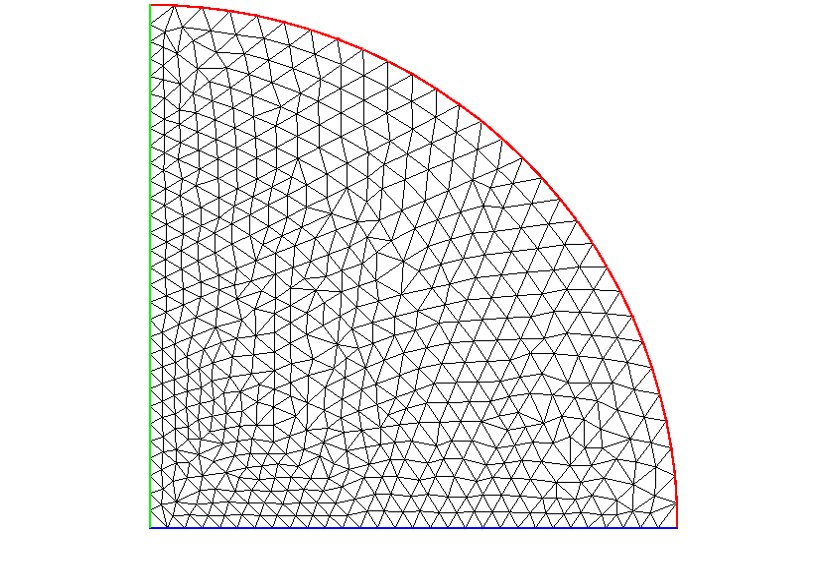}
  \hss}
\caption{
(Color online) -- Triangulation over the computational domain.
}
\label{Fig:Domain}
\end{figure}

The energy-momentum tensor of Witten's model is
\Align{Energy-Momentum}{
T^\mu_{~\nu} &= -\sum_a F^{\oa\,\mu\rho}F^\oa_{\nu\rho} + 
\sum_a(D^\mu\Phi_a)^*D_\nu\Phi_a \nonumber \\
&~~~~~+\sum_a(D_\nu\Phi_a)^*D^\mu\Phi_a-\delta^\mu_\nu \Lagrangian\,,
}
and the energy $E=\int T_t^t\dd^3x$.
Using the axially symmetric ansatz in the radial gauge,
\Align{fields2}{
A^\oa_\mu\dd x^\mu &= V^\oa \dd t + W^\oa \dd\varphi + Z^\oa \dd z\,,
 \\
   \Phi_1 &= X_1+iY_1\,, 
   \Phi_2 = (X_2+iY_2)\exp\{i(\omega t+m\varphi) \}\,, \notag 
}
where $X_a$, $Y_a$, $V^\oa$, $W^\oa$, $Z^\oa$ 
depend on $\rho,z$, gives 
\begin{widetext}
\Align{Energy-app}{
   E &=\int \left\lbrace \frac{1}{2}\sum_a\left( \grad V^{\oa\,2} 
+\frac{1}{\rho^2}\grad W^{\oa\,2}
+ \partial_\rho Z^{\oa\,2} \right)
+2\sum_a\ga Z^\oa\left(Y_a\partial_zX_a-X_a\partial_zY_a	\right)
+\grad X_1^2 + \grad Y_1^2 + \grad X_2^2 + \grad Y_2^2\right.\nonumber \\
&+\left( \left(g_1 V^\oo\right)^2 + \left(\frac{g_1}{\rho} W^\oo\right)^2 
+ \left(g_1 Z^\oo\right)^2\right)(X_1^2+Y_1^2)
+\frac{\lambda_1}{4}\left( X_1^2+Y_1^2-\eta^2_1\right)^2
+\gamma(X_1^2+Y_1^2) (X_2^2+Y_2^2)	\nonumber \\
&\left.+\left( \left(\omega-g_2 V^\ot\right)^2 
+ \frac{1}{\rho^2}\left(m-g_2 W^\ot\right)^2 
+ \left(g_2 Z^\ot\right)^2 
+\frac{\lambda_2}{4}\left( X_2^2+Y_2^2-2\eta^2_2\right) \right)(X_2^2+Y_2^2)
	\right\rbrace	\dd^3x\,,
}
\end{widetext}
where $\grad\equiv (\partial_\rho,\partial_z)$ denotes the 
gradient operator in the $(\rho,z)$-plane. Since the integrand 
in \Eqref{Energy-app} does not depend on $\varphi$ and is 
invariant under $z\to -z$, we choose the integration domain 
to be a quadrant of radius $\rmax$,
\Equation{Domain-app}{
\Omega~:=~\left\lbrace\rho, z\in[0,\rmax]\times[0,\rmax]
~\Big|~\sqrt{\rho^2+z^2}\leq\rmax\right\rbrace\,,
}
this is illustrated in \Figref{Fig:Domain}. Regularity conditions 
at the symmetry axis  require that at $\rho=0$ one has
\Align{Regularity}{
 & X_2=Y_2 =0\,,~\partial_\rho X_1=\partial_\rho Y_1=0\,,\nonumber \\
 &W^\oa=0\,,~\partial_\rho V^\oa =\partial_\rho Z^\oa =0\,.
}
The reflexion symmetry in the equatorial plane implies that 
at $z=0$ 
\Align{Reflect}{
 &  \partial_z X_a =0\,,~Y_a=0\,,\nonumber \\
&\partial_z V^\oa =\partial_z W^\oa=\partial_z Z^\oa =0\,.
}
Since the massive fields approach their asymptotic values 
exponentially fast, we assume that for $\sqrt{\rho^2+z^2}=\rmax$ 
the fields take on the vacuum values 
\Equation{Asymp-scal-app}{
   X_1=1\,,~Y_1=X_2=Y_2=0\,,~~~\text{and}~~~A^\oa_\mu=0,
}
but then we refine this by imposing mixed boundary conditions for 
the massless $A^{(2)}_\mu$ to take into account its slow falloff.  

We find stationary solutions by minimizing the energy for 
a fixed Noether charge $Q$. To this end, we express the 
frequency $\omega$ in terms of $Q$ via rewriting  \Eqref{Q} 
of the main text, in the form
\Equation{QQ}{
Q=2(\omega\Sigma_1-\Sigma_2)
}
with 
\Align{Sigma-app}{
   \Sigma_1&=\int \dd^3x\  \left(X_2^2+Y_2^2\right)\,,	\nonumber\\
   \Sigma_2&=\int \dd^3x\  g_2 V^\ot\left(X_2^2+Y_2^2\right)\,,
}
so that 
\Equation{omega}{
\omega=\frac{Q+2\Sigma_2}{2\Sigma_1}\,,
}
which should be inserted to \Eqref{Energy-app}. The energy 
is then minimized at a fixed  $Q$.

Having fixed $Q$, one has to fix also the winding numbers 
$n,m$. The number $m$ determines the windings of the phase 
of $\Phi_2\sim e^{im\varphi}$ along the azimuthal direction, 
and, after separating the angular variable, is explicitly 
contained in the integrand in \eqref{Energy-app}. The number 
$n$ determines the increase of phase of  $\Phi_1=X_1+iY_1$ 
around the boundary of $(\rho,z)$ half-plane. To fix it,
one choose $X_1(\rho,z)$ and $Y_1(\rho,z)$ which vanish, 
respectively, $n$ and $n-1$ times along the positive $z$-axis, 
for $\rho=0$ and $0<z<\infty$  \cite{Radu.Volkov:08}. This 
guarantees that $\Phi_1$ vanishes at a point $(\rho_0,0)$ 
corresponding to the center of the closed vortex forming 
the vorton, and the phase of $\Phi_1$ winds around this point. 
It is sufficient to enforce the value of $n$ only for the 
input $X_1,Y_1$ field configuration, since  later it will 
be automatically preserved in the energy minimizing iterations 
(unless the vortex center hits the $z$-axis), because `unwinding' 
would require an infinite amount of energy.    

After fixing $Q,m$ in \Eqref{Energy-app} and choosing an 
input configuration $X_1(\rho,z)$, $Y_1(\rho,z)$, $X_2(\rho,z)$ 
with the prescribed value of $n$, the iterative energy minimization 
produces a sequence of regular, vorton-type configurations. However, 
these do not necessarily converge to a smooth limiting configuration,
even though the energy is bounded from below. If $Q$ is small, 
then the values of $\rho_0$ typically converge to zero and the 
vorton loops shrink. Physically, this means that there are no 
enough field quanta to condense and make a vorton. There are no 
stationary solutions in this case and the stationary energy minimizer 
is trying to approximate a non-stationary collection of non-condensed 
quanta. On the other hand, for a large enough $Q$ the $\rho_0$-values 
converge to a finite limit corresponding to the radius of the stationary 
vorton.

The variational problem of minimizing \Eqref{Energy-app} is 
defined for numerical computations by adapting a finite element 
formulation provided by the Freefem++ library \cite{Freefem}. 
The discretization within this formulation is done via a 
(homogeneous) triangulation over $\Omega$, based on the 
Delaunay-Voronoi algorithm. The functions are expanded  with 
respect to a continuous piecewise quadratic polynomial basis 
on each triangle. 
The accuracy of the method is controlled by the number of 
triangles, by the order of expansion in the basis for each 
triangle, and also by the order of the quadrature formula 
for the integral over the triangles. 
Once the problem is mathematically posed, a numerical optimization 
algorithm is used to solve the nonlinear variational problem of 
finding the minima of $E$. We used a nonlinear conjugate gradient 
method. The algorithm is iterated until the relative variation of 
the norm of the gradient of the functional $E$ with respect to 
all degrees of freedom is less than $10^{-6}$. Virial relations 
which should be fulfilled by the solutions are then checked.

\subsection{Q-balls versus Vortons}

One can consistently set to zero in the field equations the 
imaginary part of the first scalar and the first gauge field, 
$Y_1=\Im (\Phi_1)=0$, $A^{(1)}_\mu=0$. Solutions obtained in 
this case are similar to the gauged Q-balls described in 
\cite{Lee.Stein-Schabes.ea:89}. $\Phi_1$ is then real and 
has no phase winding ($n=0$), while $\Phi_2\sim e^{im\varphi}$ 
remains complex-valued. Q-balls with $m=0$ are spherically 
symmetric and non-spinning, while those with $m\neq 0$ are 
axially symmetric and support a non-zero angular momentum $J=mQ$. 
Spinning Q-balls have toroidal profiles similar to those for 
vortons. For Q-balls the field $\Phi_1$ is small inside the 
torus tube, although it does not vanish at the center of the 
tube as for vortons. Since $\Phi_1$ is real, the first Noether 
current $j^\mu_1$ vanishes, and there is no flux of 
$\vec{B}^{(1)}$ along the torus tube. However, for Q-balls as 
for vortons, the phase of $\Phi_2$  winds {along} the torus 
tube giving rise to an azimuthal current $j^\mu_2$ and to a 
dipole magnetic field $\vec{B}^{(2)}$. Both Q-balls and vortons 
carry an electric charge 
${\cal Q}=\oint \dd\vec{E}^{(2)}\cdot \dd\vec{S}=g_2Q$. 
\begin{figure}[!htb]
\hbox to \linewidth{ \hss

\begingroup
  \makeatletter
  \providecommand\color[2][]{%
    \GenericError{(gnuplot) \space\space\space\@spaces}{%
      Package color not loaded in conjunction with
      terminal option `colourtext'%
    }{See the gnuplot documentation for explanation.%
    }{Either use 'blacktext' in gnuplot or load the package
      color.sty in LaTeX.}%
    \renewcommand\color[2][]{}%
  }%
  \providecommand\includegraphics[2][]{%
    \GenericError{(gnuplot) \space\space\space\@spaces}{%
      Package graphicx or graphics not loaded%
    }{See the gnuplot documentation for explanation.%
    }{The gnuplot epslatex terminal needs graphicx.sty or graphics.sty.}%
    \renewcommand\includegraphics[2][]{}%
  }%
  \providecommand\rotatebox[2]{#2}%
  \@ifundefined{ifGPcolor}{%
    \newif\ifGPcolor
    \GPcolorfalse
  }{}%
  \@ifundefined{ifGPblacktext}{%
    \newif\ifGPblacktext
    \GPblacktexttrue
  }{}%
  \let\gplgaddtomacro\g@addto@macro
  \gdef\gplbacktext{}%
  \gdef\gplfronttext{}%
  \makeatother
  \ifGPblacktext
    \def\colorrgb#1{}%
    \def\colorgray#1{}%
  \else
    \ifGPcolor
      \def\colorrgb#1{\color[rgb]{#1}}%
      \def\colorgray#1{\color[gray]{#1}}%
      \expandafter\def\csname LTw\endcsname{\color{white}}%
      \expandafter\def\csname LTb\endcsname{\color{black}}%
      \expandafter\def\csname LTa\endcsname{\color{black}}%
      \expandafter\def\csname LT0\endcsname{\color[rgb]{1,0,0}}%
      \expandafter\def\csname LT1\endcsname{\color[rgb]{0,1,0}}%
      \expandafter\def\csname LT2\endcsname{\color[rgb]{0,0,1}}%
      \expandafter\def\csname LT3\endcsname{\color[rgb]{1,0,1}}%
      \expandafter\def\csname LT4\endcsname{\color[rgb]{0,1,1}}%
      \expandafter\def\csname LT5\endcsname{\color[rgb]{1,1,0}}%
      \expandafter\def\csname LT6\endcsname{\color[rgb]{0,0,0}}%
      \expandafter\def\csname LT7\endcsname{\color[rgb]{1,0.3,0}}%
      \expandafter\def\csname LT8\endcsname{\color[rgb]{0.5,0.5,0.5}}%
    \else
      \def\colorrgb#1{\color{black}}%
      \def\colorgray#1{\color[gray]{#1}}%
      \expandafter\def\csname LTw\endcsname{\color{white}}%
      \expandafter\def\csname LTb\endcsname{\color{black}}%
      \expandafter\def\csname LTa\endcsname{\color{black}}%
      \expandafter\def\csname LT0\endcsname{\color{black}}%
      \expandafter\def\csname LT1\endcsname{\color{black}}%
      \expandafter\def\csname LT2\endcsname{\color{black}}%
      \expandafter\def\csname LT3\endcsname{\color{black}}%
      \expandafter\def\csname LT4\endcsname{\color{black}}%
      \expandafter\def\csname LT5\endcsname{\color{black}}%
      \expandafter\def\csname LT6\endcsname{\color{black}}%
      \expandafter\def\csname LT7\endcsname{\color{black}}%
      \expandafter\def\csname LT8\endcsname{\color{black}}%
    \fi
  \fi
  \setlength{\unitlength}{0.0500bp}%
 \resizebox{\linewidth}{!}{ \begin{picture}(7200.00,5040.00)%
    \gplgaddtomacro\gplbacktext{%
      \csname LTb\endcsname%
      \put(1078,704){\makebox(0,0)[r]{\strut{} 0.75}}%
      \put(1078,1286){\makebox(0,0)[r]{\strut{} 0.8}}%
      \put(1078,1867){\makebox(0,0)[r]{\strut{} 0.85}}%
      \put(1078,2449){\makebox(0,0)[r]{\strut{} 0.9}}%
      \put(1078,3030){\makebox(0,0)[r]{\strut{} 0.95}}%
      \put(1078,3612){\makebox(0,0)[r]{\strut{} 1}}%
      \put(1078,4193){\makebox(0,0)[r]{\strut{} 1.05}}%
      \put(1078,4775){\makebox(0,0)[r]{\strut{} 1.1}}%
      \put(1210,484){\makebox(0,0){\strut{} 2000}}%
      \put(2329,484){\makebox(0,0){\strut{} 4000}}%
      \put(3447,484){\makebox(0,0){\strut{} 6000}}%
      \put(4566,484){\makebox(0,0){\strut{} 8000}}%
      \put(5684,484){\makebox(0,0){\strut{} 10000}}%
      \put(6803,484){\makebox(0,0){\strut{} 12000}}%
      \put(176,2739){\rotatebox{-270}{\makebox(0,0){\strut{}E/Q}}}%
      \put(4006,154){\makebox(0,0){\strut{}Q}}%
    }%
    \gplgaddtomacro\gplfronttext{%
      \csname LTb\endcsname%
      \put(3517,1317){\makebox(0,0)[r]{\strut{}$m=0$ Q-Ball}}%
      \csname LTb\endcsname%
      \put(3517,1097){\makebox(0,0)[r]{\strut{}$m=1$ Q-ball}}%
      \csname LTb\endcsname%
      \put(3517,877){\makebox(0,0)[r]{\strut{}$m=1$ Vorton}}%
    }%
    \gplgaddtomacro\gplbacktext{%
      \csname LTb\endcsname%
      \put(4230,2708){\makebox(0,0)[r]{\strut{} 0.855}}%
      \put(4230,3616){\makebox(0,0)[r]{\strut{} 0.86}}%
      \put(4230,4523){\makebox(0,0)[r]{\strut{} 0.865}}%
      \put(4362,2488){\makebox(0,0){\strut{} 8000}}%
      \put(5403,2488){\makebox(0,0){\strut{} 8200}}%
      \put(6443,2488){\makebox(0,0){\strut{} 8400}}%
    }%
    \gplgaddtomacro\gplfronttext{%
    }%
    \gplbacktext
    \put(0,0){\includegraphics{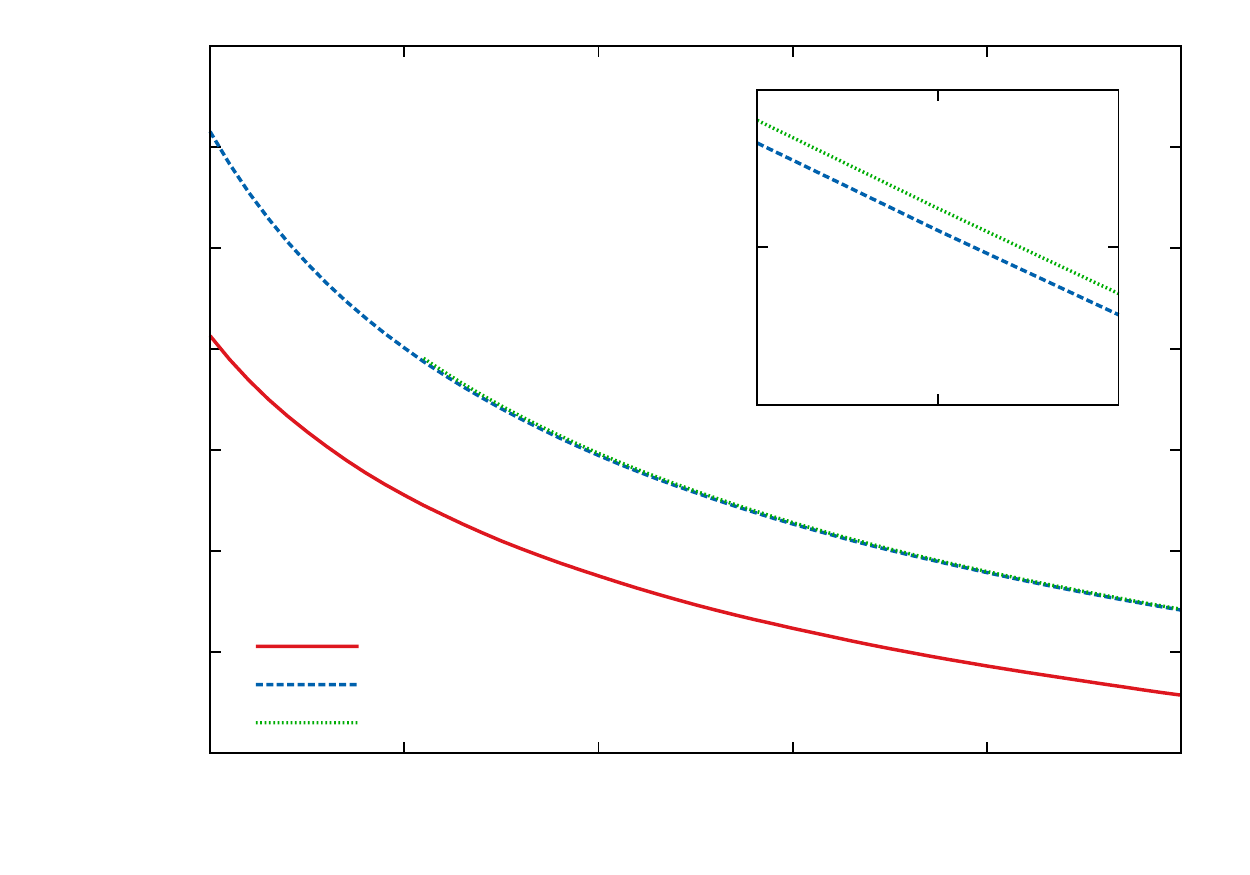}}%
    \gplfronttext
  \end{picture}%
  }
\endgroup

\hss}
\caption{
(Color online) -- 
The energy-to-charge ratio $E/Q$ against $Q$ for Q-balls and vortons 
for the same parameters as in Fig.~1 in the main text. The lower 
(red) curve corresponds to the $m=0$ Q-balls. The upper (blue-green) 
curve corresponds to the $m=1$ Q-balls and vortons whose energies turn 
out to be almost the same. Zooming this curve shown in the insertion 
reveals that the vorton energy (dotted green line) is slightly larger 
than the Q-ball energy (dashed blue line). 
}
\label{Fig:Qball}
\end{figure}
For a given (large enough) charge $Q$ and an azimuthal winding 
number $m$, Q-balls  and vortons have the same angular momentum 
$J=mQ$. However, their energies $E$ are not the same,  Q-balls 
being less energetic  (see \Figref{Fig:Qball}). Therefore, vortons 
are not energetically protected from decaying into Q-balls,
although they cannot decay into free particles (since $E/Q<1$). 

The lightest solution for a given $Q$ is the spherically symmetric 
Q-ball with $J=0$, but vortons cannot decay to it due to the 
angular momentum conservation, unless they break in several
components spiralling around the total center of mass. 
It seems this is exactly what happens for the $m>2$ vortons,
as described in the main text. Specifically, they seem to unwind 
to $n=0$ via shrinking the $\Phi_1=0$ line to zero and split into 
small spherical Q-balls whose total orbital angular momentum 
equals the initial $J$.   

However,  the $m=1$ vorton is `hard to pinch' and it does not 
break into non-spinning Q-balls.  On the other hand, it could 
in principle decay into the $m=1$ spinning Q-ball with the same 
$Q,J$, although their relative energy difference is only $\sim 10^{-3}$ 
(see \Figref{Fig:Qball}). To unwind to the Q-ball, the vorton has 
to get rid of the phase winding around the closed line where 
$\Phi_1=0$. This line cannot break, since this would cost an
infinite energy, but it can shrink to zero at a finite energy 
cost. To estimate the energy needed, we considered an interpolating 
sequence of trial field configurations for which $A^{(a)}_\mu$ and 
$\Phi_2$ are the same as for the $m=1$ stationary vorton, while 
$\Phi_1(x^\mu)$ is replaced by $\Phi_1(\lambda x^\mu)$ with $\lambda$ 
being the scale parameter. We then calculate the total energy 
$E(\lambda)$, which coincides with the vorton energy for $\lambda=1$. 
The $\Phi_1=0$ line shrinks to zero when $\lambda\to \infty$, 
and it turns out that $E(\infty)\approx 3\times E(1)$, which gives 
an idea of the energy needed to unwind. Therefore, $m=1$ vortons 
and Q-balls are separated  by a potential barrier whose hight is 
of the same order as the vorton energy. Since vortons are rather 
heavy, a very strong perturbation is needed to unwind them into 
Q-balls. As mentioned in the main text, the vorton unwinding 
could also be disfavoured by quantum effects. 

The conclusion is that the $m=1$ vortons are dynamically stable 
and could, perhaps, be destroyed only by very strong perturbations.

\subsection{Hyperbolic evolution}

To investigate the dynamical stability of the global vortons, 
the axially symmetric solutions  are used as the initial data 
for a $3+1$ evolution code. This time the integration domain 
$\Omega$ is an (open) bounded subset of $\Real^3$ with the 
boundary $\partial\Omega$ and with the attached normal derivative 
$\partial_n$.  We choose $\Omega$ to be the interior of a sphere 
of radius $\rmax$. The discrete version is obtained by a homogeneous 
triangulation over $\Omega$.
The following numerical scheme was used to evolve the vortons. 
The dynamical equations for the scalars read in the global limit  
\Align{Dynamical}{
   \partial_{tt} &\Phi_a-\Delta\Phi_a +\frac{\partial V}{\partial |\Phi_a|^2}\Phi_a=0	\,.
}
Here $\Delta\equiv\vec{\nabla}^2$ is the Laplace operator and 
$\vec{\nabla}$ the gradient operator in three space dimensions 
in Cartesian coordinates. The weak formulation of the hyperbolic 
problem \Eqref{Dynamical} is obtained by multiplying the equation 
by test functions $w_a$, integrating over $\Omega$ and applying 
the Stokes formula,  
\Align{DynamicalWeak1}{
   \int_\Omega w_a\partial_{tt}& \Phi_a 
+\int_\Omega \vec{\nabla} w_a\cdot\vec{\nabla}\Phi_a 
+ \int_\Omega w_a\frac{\partial V}{\partial |\Phi_a|^2}\Phi_a	\nonumber \\
&~~~~~~ -\int_{\partial\Omega} w_a\partial_n\Phi_a=0	\,.
}
The time discretization is achieved using a $\beta$-Newmark scheme, 
which converts \Eqref{DynamicalWeak1} to  
\Align{DynamicalWeakDiscrete}{
     \int_\Omega w_a& \frac{\Phi_a^{[n+1]}-2\Phi_a^{[n]}
+\Phi_a^{[n-1]}}{\dd t^2} 		\nonumber \\
      &+\int_\Omega\vec{\nabla} w_a\cdot\vec{\nabla}
\left(\beta\Phi_a^{[n+1]}+(1-2\beta)\Phi_a^{[n]}+\beta\Phi_a^{[n-1]} \right) \nonumber \\
      &+\int_\Omega w_a\frac{\partial V^{[n]}}{\partial |\Phi_a|^2}\Phi_a^{[n]}
	    -\int_{\partial\Omega} w_a\partial_n\Phi_a^{[n]}=0	\,,
}
where $\Phi_a^{[n]},V^{[n]}$ are the value of $\Phi_a,V$ at the 
time moment $t_0+ndt$. Here $0\leq\beta\leq1$ and the scheme is 
unconditionally stable for $\beta\geq1/4$. Typically we choose 
$\beta=1/4$, which corresponds to the constant average acceleration 
method, while choosing $\beta=1/2$ would reproduce the Crank-Nicholson 
scheme. We assume that $\partial_t\Phi_a=0$ at $\partial\Omega$, 
so that the boundary values of fields are frozen.

The time-discretized equations \Eqref{DynamicalWeakDiscrete} can 
be rewritten as a recurrence 
\Equation{Recurrence}{
   u^{[n+1]}=\bs A^{-1}\left(\bs B u^{[n]}+ \bs C	\right)-u^{[n-1]}\,
}
where $u^{[n]}\equiv \{u^{[n]}_i\}$ 
is the vector containing all discretized degrees of freedom at 
the $n$-th time step. $\bs A$, $\bs B$ are matrix bilinear operators 
and $\bs C$ is a vector containing the nonlinearities,
\Align{Matrices}{
\bs A_{ij}&=	\int_\Omega (u_j^{[n]}u_i^{[n]}
   +\beta_1\vec{\nabla} u_j^{[n]}\cdot\vec{\nabla} u_i^{[n]})
    -\int_{\partial\Omega}\beta_1 u_j^{[n]}\partial_n u_i^{[n]}		\, , \nonumber \\
\bs B_{ij}&=	\int_\Omega (2u_j^{[n]}u_i^{[n]}
   -\beta_2\vec{\nabla} u_j^{[n]}\cdot\vec{\nabla} u_i^{[n]})
    +\int_{\partial\Omega}\beta_2 u_j^{[n]}\partial_n u_i^{[n]}		\, , \nonumber \\
\bs C_i&= -\dd t^2\int_\Omega 	
u^{[n]}_i\frac{\partial V^{[n]}}{\partial |\Phi_a|^2}\Phi_a^{[n]}\,,		\, 
}
with $\beta_1=\beta\dd t^2$ and $\beta_2=(1-2\beta)\dd t^2$.
Using the stationary, axially symmetric solutions obtained by 
minimizing the energy, the recurrence is initialized by 
\Align{RecurrenceInit}{
   \Phi_1^0&=	X_1(\rho,|z|)+i\sg(z)Y_1(\rho,|z|)	\,,&\Phi_1^1&=	 \Phi_1^0\nonumber\\ 
  \Phi_2^0&=	(X_2(\rho,|z|)+i\sg(z)Y_2(\rho,|z|))\Exp{i m\varphi}	\,,&\Phi_2^1&=\Phi_2^0\Exp{i \omega \dd t}.
}
The time step $\dd t$ is typically chosen to be $0.1$ and the 
configuration is evolved for several hundreds of internal periods 
$T=2\pi/\omega$.

It should be stressed that the passage from the stationary, 
axially symmetric vortons to their dynamical counterparts starts 
by uplifting the former from two to three spatial dimensions. 
In doing this, the stationary solutions obtained on the 2D mesh 
of the type shown in \Figref{Fig:Domain} are interpolated to adapt 
them to a 3D mesh built  via a 3D triangulation over the 3D 
integration region. The two meshes are completely different and 
have neither common symmetries nor shared vertices.  The uplifted 
to 3D field configurations thus differ from the very beginning from 
the true 3D stationary solutions. They correspond instead to perturbed 
stationary solutions, and it is this initial perturbation which 
triggers a non-trivial temporal dynamics. The perturbation 
turns out  to be  strong enough to destroy the  $m>2$ vortons 
almost immediately. Therefore, since the $m=1$ vortons survive, 
this strongly suggests that they are dynamically stable.

\end{document}